# Compliance Costs of AI Technology Commercialization: A Field Deployment Perspective

Weiyue Wu and Shaoshan Liu

## 1. Introduction

While Artificial Intelligence (AI) technologies are progressing fast, compliance costs have become a huge financial burden for AI startups, which are already constrained on research & development budgets. This situation creates a compliance trap, as many AI startups are not financially prepared to cope with a broad spectrum of regulatory requirements. Particularly, the complex and varying regulatory processes across the globe subtly give advantages to well-established and resourceful technology firms over resource-constrained AI startups [1]. The continuation of this trend may phase out the majority of AI startups and lead to giant technology firms' monopolies of AI technologies. To demonstrate the reality of the compliance trap, from a field deployment perspective, we delve into the details of compliance costs of AI commercial operations.

## 2. Financial Vulnerability: Tech Giants vs. AI Startups

Compared to established tech giants, AI startups are much more financially vulnerable. Based on the OECD Regulatory Compliance Cost Assessment Guidance [2], we quantitatively compare the financial vulnerability of tech giants versus AI startups. Conducting a financial statement simulation provides a glimpse at the impact of changes in compliance costs. The simulation in Table 1 assumes that gross profit and expense are a fixed percentage of revenue. The compliance cost is split into a fixed cost regardless of revenue and a variable cost that takes 5% of revenue. When the fixed compliance cost increases by 200%, the operating margin of the startup changes from 13% to -7%, turning the company from a profitable business into a money-losing one. By contrast, such a change only brings a slight drop in operating margin in tech giants.

|  | Start-ups | | Tech Giants | |
|---|---|---|---|---|
|  | Base case | 200% increase in fixed compliance cost | Base case | 200% increase in fixed compliance cost |
| Revenue | $2,000,000 | $2,000,000 | $20,000,000 | $20,000,000 |
| Gross Margin | 40% | 40% | 40% | 40% |
| Gross Profit | 800,000 | 800,000 | 8,000,000 | 8,000,000 |
| Total Compliance Cost | 300,000 | 700,000 | 1,200,000 | 1,600,000 |
|    Fixed compliance cost | 200,000 | 600,000 | 200,000 | 600,000 |
|    Variable compliance cost | 100,000 | 100,000 | 1,000,000 | 1,000,000 |
| Expenses | 240,000 | 240,000 | 2,400,000 | 2,400,000 |
| Operating Income | 260,000 | (140,000) | 4,400,000 | 4,000,000 |
| **Operating Margin** | **13%** | **-7%** | **22%** | **20%** |

Table 1: Sensitivity Analysis: Impact of Change in Compliance Cost on Operating Income

The actual situation of AI startups is more complex than this estimation, as it is nearly impossible for external analysts to estimate the compliance costs. According to Accounting Standards, companies are not required to disclose compliance cost explicitly in financial reports when the cost is not material [3]. Thus, compliance costs may become hidden by nature and classified into other categories, such as Research and Development expenses and other administrative expenses. Lacking first-hand information, analysts on the macro-economic level tend to underestimate the costs of AI regulations. For instance, in an impact assessment report of the Europe AI Act, the estimated annual compliance cost of one AI product that averagely costs EUR 170,000 to develop is EUR 29,277, we believe this study has underestimated the actual costs of AI compliance [4].

## 3. The Compliance Trap

AI is a highly regulated industry, but unfortunately, there is no standardized AI regulation framework, and hence compliance costs often become a financial trap for AI startups [1]. Most AI entrepreneurs may not even be aware of the existence of compliance costs, let alone the severe impact compliance costs may have on the company's overall financial health. As illustrated in Figure 1, we summarize the following challenges.

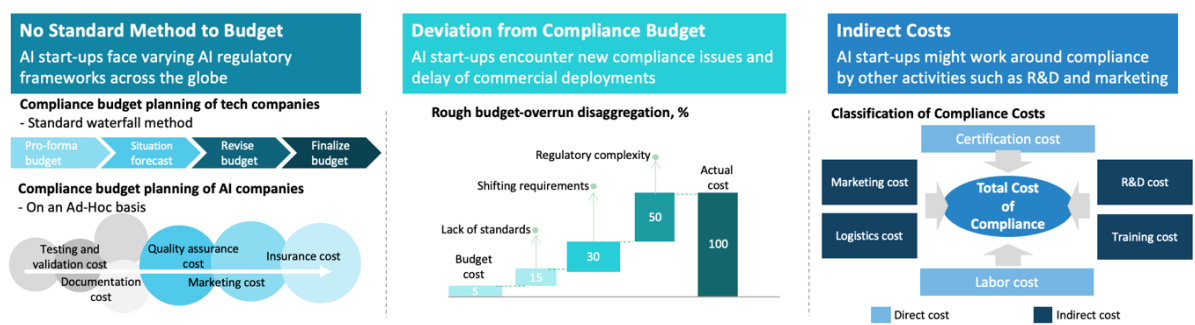

Figure 1: the compliance trap for AI startups

First, unlike R&D budgeting, due to varying AI regulatory frameworks across the globe or even across multiple regions within a country, there is **no standard method to budget** for AI compliance costs. Even estimating the range of AI compliance costs is infeasible.

Second, even with an AI compliance budget, the actual costs may **significantly deviate from the budget**. AI startups often encounter new compliance issues as they progress through commercialization. In addition, opportunity costs arise as regulators inspect AI products on safety and privacy issues, causing delays in commercial deployments.

Third, varying AI regulations often introduce **indirect costs**. For instance, a strict compliance environment demands engineers deal with regulatory issues such as responding to various compliance technical inquiries instead of spending time developing products. Such a shift of focus does not reflect in financial reports, as engineers' costs are categorized as R&D costs.

## 4. A Field Deployment Perspective

In this section, with more than six years of first-hand experience in deploying commercial autonomous driving services, we delve into the details of compliance costs from a field deployment perspective, in the hope that the insights we provide can raise awareness of the adverse impact of the lack of standardized AI regulations.

### 4.1. Background

PerceptIn is an autonomous driving startup founded in 2016. It offers autonomous micro-mobility solutions to customers from the United States, European Union, and Asia. The company only budgeted for ordinary compliance expenses, such as the direct labor cost of a safety driver on board and the equipment cost of a waterproof surveillance camera. While facing a broad spectrum of regulatory obstacles across different countries, PerceptIn had fallen into the compliance trap. Many financial and human resources have been spent out of budget to comply with various regulatory frameworks in different regions.

### 4.2. Scenario 1 – No Standard Method to Budget

The AI regulation framework in China was blurry, and when the company first launched the autonomous micro-mobility project in China, it was impossible to budget for compliance costs. For instance, since relevant regulations were absent back then, the company needed to develop its own testing plan to obtain deployment approval. Without detailed testing standards, the company had to spend $25,000 per month to simulate real-world scenarios at the initial stage for a testing site for testing and demonstration purposes. The testing process is to obtain detailed validation results, whereas the demonstration process is to convince the regulatory body regarding the safety and reliability of the service. The $300,000 annual cost was not included in the company's original budget and imposed a heavy burden on the company's financial health.

### 4.3. Scenario 2 – Deviation from Compliance Budget

The company was invited to launch an autonomous driving pilot program in a European city. Before rolling out the project, the company was asked to prepare a risk mitigation plan for 40 different scenarios. To cope with the regulatory process, the R&D team shifted its focus to responding to scenarios-based functional specifications and supplemented the mitigation plan with real-time data. During the project budgeting phase, the company had prepared 20 man-days to cope with the AI regulatory process. Nevertheless, the process turned out to consume 400 man-days to complete the process. While the original budget was $10,000, in the end, the process consumed $200,000. Such a severe mismatch was caused by the lack of a standardized regulatory process, as any response from our technical team would bring on the next round of regulatory questions.

### 4.4. Scenario 3 – Indirect Costs

Japan is famous for its rigid structure in organizations. Thus without an established compliance process in place, gaining the confidence of the Japanese government is essential to commercial deployment. To gain the confidence of the Japanese government, the company first debuted a marketing campaign to promote safe autonomous micro-mobility services in a smart city project [5]. With a successful local case and globally established brand, the company then discussed operation permits with the Ministry of Land, Infrastructure, Transport, and Tourism (MLIT) [6]. The preparation and initiation of the project took over 24 months, costing $500,000 in promotion, material preparation, and marketing campaigns. Traditionally marketing activities were not meant to cope with compliance requirements. However, in this case, marketing was a tool to convince the regulatory body to further autonomous driving operation permits.

In the case of PerceptIn, the compliance cost of one deployment project is $ 344,000 on average, whereas the average R&D cost is around $150,000, making the **compliance costs 2.3 times the amount of R&D costs, far exceeding the 17.6% estimation of the Europe AI Act.**

## 5. The Silver Linings

The root cause of the compliance trap is the lack of a standardized AI regulatory framework. An ultimate solution to this problem lies in creating a global golden standard for AI regulation. A study of the Food and Drug Administration's (FDA's) history sheds light on properly regulating a new field. First, pharmaceutical products a century ago and AI today are both viewed as black boxes. Even the most sophisticated scientists could not predict their development trajectory, let alone legal experts. Second, pharmaceutical and AI technologies can potentially cause severe public risks if they are not adequately regulated. Third, both industries have enormous potential for improving people's well-being. Like the FDA, a consumer protection agency shall be established to ensure that AI is developed for people's well-being. Such an agency should develop the expertise and capability to scientifically judge whether an AI product is ethical and legal. Such an agency should provide guidance to various governments within the U.S. and worldwide on AI regulations.

However, before a consensus can be reached regarding the golden standard, a new business model, Compliance-as-a-Service (CaaS), can specialize in dealing with varying AI regulatory frameworks and thus amortize compliance costs across different AI startups. In addition, CaaS reduces the friction between regulatory bodies and AI startups by providing an interface to compile legal terms into technical and operational plans. With the new business model, AI entrepreneurs can adequately budget for compliance when evaluating the potential of an innovative idea.

## 6. Summary

AI is a promising industry mainly filled with startups exploring the applications of AI technologies in different aspects of our daily lives. Compared to well-established tech giants, AI startups are financially vulnerable. Unfortunately, the lack of standardized AI regulatory frameworks creates a compliance trap that may destroy an AI startup financially, which could lead to a more profound impact of creating a competitive advantage for tech giants over AI startups. We have examined the details of compliance costs from a field deployment perspective to demonstrate the reality of the compliance trap. Ideally, if a global golden standard on AI regulation could be developed, then AI startups could accurately budget for compliance costs. However, before a consensus can be reached regarding the golden standard, we believe that a new business model, compliance as a service, can specialize in dealing with varying AI regulatory frameworks and thus amortize compliance costs across different AI startups.

## Biography:

**Weiyue Wu** is Chief Operating Officer of PerceptIn, an autonomous driving startup founded in 2016. At PerceptIn, she has been in charge of commercial autonomous driving service deployments in the US, Europe, Japan, and China. Before PerceptIn, she served as Investment Director of Oxford Seed Fund and Investment Advisor of ARM Accelerator. She began her career as a Multi-National Corporation Compliance Auditor at KPMG and a Senior Automobile Consultant at Deloitte. She received her MBA from the University of Oxford. She is a founding member of IEEE Special Technical Community on Autonomous Driving Technologies, a Certified Public Accountant and a practicing lawyer in China.

**Dr. Shaoshan Liu**'s background is a unique combination of technology, entrepreneurship, and public policy, which enables him to take on great global challenges. On technology, Dr. Liu has published 4 textbooks, more than 100 research papers, and holds more than 150 patents in autonomous systems. On entrepreneurship, Dr. Shaoshan Liu is CEO of PerceptIn and has commercially deployed autonomous micro-mobility services in the U.S., Europe, Japan, and China *etc.* He is the Asia Chair of IEEE Entrepreneurship. On public policy, Dr. Liu has served on the World Economic Forum's panel on Industry Response to Government Procurement Policy, is leading the Autonomous Machine Computing roadmap under IEEE International Roadmap of Devices and Systems (IRDS) and is a member of the ACM U.S. Technology Policy Committee. Dr. Liu's educational background includes a M.S. in Biomedical Engineering, a Ph.D. in Computer Engineering from the U.C. Irvine, and a Master of Public Administration (MPA) from Harvard University. He is an IEEE Senior Member, an IEEE Computer Society Distinguished Speaker, an ACM Distinguished Speaker, an Advisory Council member of

Harvard Business Review, a member of MIT Technology Review's Global Insights Panel, and a member of the Forbes Technology Council.